\documentclass[fleqn,usenatbib]{mnras}

\usepackage{newtxtext,newtxmath}
\usepackage[T1]{fontenc}
\usepackage{ae,aecompl}

\usepackage{graphicx}	% Including figure files
\usepackage{dblfloatfix}
\usepackage{amsmath}	% Advanced maths commands
\usepackage{threeparttable}
\usepackage{subcaption}
\usepackage{caption}
\usepackage{color}
\usepackage[dvipsnames]{xcolor}
\usepackage[normalem]{ulem}
\usepackage{booktabs}
\usepackage{multirow}

\usepackage{mathtools}
\newcommand{\fermi}{\textit{Fermi}-{\rm LAT}}
\newcommand{\planck}{\textit{Planck}}

\def\deg{\hbox{$^\circ$}}

\title[Gamma-ray emission around Rosette Nebula]{Diffuse gamma-ray emission around the Rosette Nebula}

\author[Liu et.al]{Jia-hao Liu$^{1,2,3}$, 
Bing Liu$^{1,2,3}$\thanks{E-mail: lbing@ustc.edu.cn},
Rui-zhi Yang$^{1,2,3}$\\
$^{1}$Deep Space Exploration Laboratory/School of Physical Sciences, University of Science and Technology of China, Hefei 230026, China\\
$^{2}$CAS Key Laboratory for Research in Galaxies and Cosmology, Department of Astronomy, School of Physical Sciences,\\
University of Science and Technology of China, Hefei, Anhui 230026, China \\
$^{3}$School of Astronomy and Space Science, University of Science and Technology of China, Hefei, Anhui 230026, China \\
}

\pubyear{2022}

\begin{document}
\label{firstpage}
\pagerange{\pageref{firstpage}--\pageref{lastpage}}
\maketitle

\begin{abstract}
The Rosette Nebula is a young stellar cluster and molecular cloud complex, located at the edge of the southern shell of a middle-aged SNR Monoceros Loop (G205.5+0.5). We revisited the GeV gamma-ray emission towards the Rosette Nebula using more than 13 years of \fermi\ data. We tested several spatial models and found that compared to the result using the CO gas template only, the inclusion of the \ion{H}{ii} gas template can significantly improve the likelihood fit. We performed spectral analysis using the new spatial template. With both the gamma-ray observation and CO+\ion{H}{ii} gas data, we derived the cosmic ray spectrum of different components in the vicinity of the Rosette Nebula. We found the gamma-ray emissions from Rosette Nebula are substantially harder than previously reported, which may imply that Rosette Nebula is another example of a gamma-ray emitting young massive star cluster.

\end{abstract}


%%%%%%%%%%%%%%%%%%%%%%%%%%%%%%%%%%%%%%%%%%%%%%%%%%

%%%%%%%%%%%%%%%%% BODY OF PAPER %%%%%%%%%%%%%%%%%%

%%%%%%%%%%%%%%%%% introduction %%%%%%%%%%%%%%%%%%
\section{Introduction}
\label{sec:intro}

As a complex of young massive clusters (YMCs) and molecular clouds (MCs), the Rosette Nebula is regarded as an important astrophysical laboratory for the studying early evolution of stars. A giant OB association NGC 2244 is located in the center of the Rosette Nebula. It generates an expanding \ion{H}{ii} region, which interacts with the surrounding the Rosette Molecular Cloud (RMC) and produces a photon-dominated region at their interface. To the west of NGC 2244, another star cluster  NGC 2237 lies on the boundary between the \ion{H}{ii} region and the cold neutral molecular material \citep{1980Blitz,2005li,2008NGC2237}.  \cite{2009Bonatto} suggested that NGC 2237 may be a young cluster located in the background of NGC 2244. The structure of NGC 2237 may contribute to the petal-like morphology of the Rosette Nebula \citep{rosette_chandra3}.
The ionized nebula interacts with the RMC, which is located to the east of the Rosette Nebula and has a collection of embedded young stellar clusters \citep{1997clusters,wang2009}. 

Along the line of sight, the Rosette Nebula partially overlaps with the edge of the southern shell of the Monoceros Loop (G205.5+0.5). Monoceros Loop is a well-studied middle-aged supernova remnant (SNR), and has a diameter of $\sim 3.8\deg$ \citep{Monoceros}, allowing detailed morphological studies in high-energy gamma rays. By analyzing about 5.5 years of {\sl Fermi}-LAT data, \cite{Monoceros} found extended gamma-ray emissions associated with the Monoceros Loop and the Rosette Nebula, and proposed that these gamma rays have the same origin which is the interaction between the hadrons accelerated by the SNR and the interstellar gas. 
 
However, we cannot formally rule out the different origins of gamma rays in the Rosette Nebula and Monoceros SNR \citep{Bourriche2022}. Indeed, in addition to SNRs, massive stars in YMCs were also proposed as major contributors to Galactic cosmic rays (CRs) \citep{Aharonian19}.  Recently, a series of observations found gamma-ray emissions that possibly originate from the  CR nuclei accelerated in YMCs interacting with the surrounding gases, such as studies of Cygnus cocoon \citep{Ackermann11, Aharonian19}, Westerlund 1 \citep{Abramowski12}, Westerlund 2 \citep{Yang18}, NGC 3603 \citep{Yang17}, 30 Dor C \citep{Abramowski15}, RSGC 1 \citep{sunRSGC1}, W40 \citep{sunw40}, Mc20 \citep{sun22}, and NGC 6618 \citep{liub2022m17}.   Furthermore, works have been done to determine the distances of NGC 2244 and the Rosette Nebula, and the results range from 1.4 kpc to 1.7 kpc \citep{1981dis,2000dis,2002dis}. Recently, from the change of stellar extinction, \cite{2018zhao} derived the distances to the Rosette Nebula and the Monoceros SNR are 1.55 kpc and 1.98 kpc, respectively, implying no interaction between the nebula and the SNR. Thus it is also possible that the gamma rays in the Rosette Nebula originated from CRs accelerated by YMCs therein. With the accumulated \fermi\ data, we revisited the Rosette Nebula region to explore the origin of the possibly associated gamma-ray emissions to test such a hypothesis.

The paper is organized as follows. In Sec.\ref{sec:gas} we described the gas content in this region. The procedure and results of data analysis are described in Sec.\ref{sec:data}. The calculation of CR spectrum and density is described in Sec.\ref{sec:CR}. Finally, we present a discussion of our results and former works and draw our conclusions in Sec.\ref{sec:conc}.

\section{Gas content around the Rosette Nebula}

According to the previous study, the gamma rays around the nebula are generated from the hadronic interactions between CRs and interstellar gas \citep{Monoceros}. As a result, the spatial distribution of gamma rays would generally follow that of gas. Thus, we first studied the spatial distribution of the gas content around the Rosette Nebula. For the following analysis, we adopted a distance of 1.55 kpc for the Rosette Nebula and the gas content around it.

\begin{figure*}
\label{fig:gas1}
    \centering
    \begin{subfigure}[t]{0.4\linewidth}
        \centering
        \includegraphics[width=\linewidth]{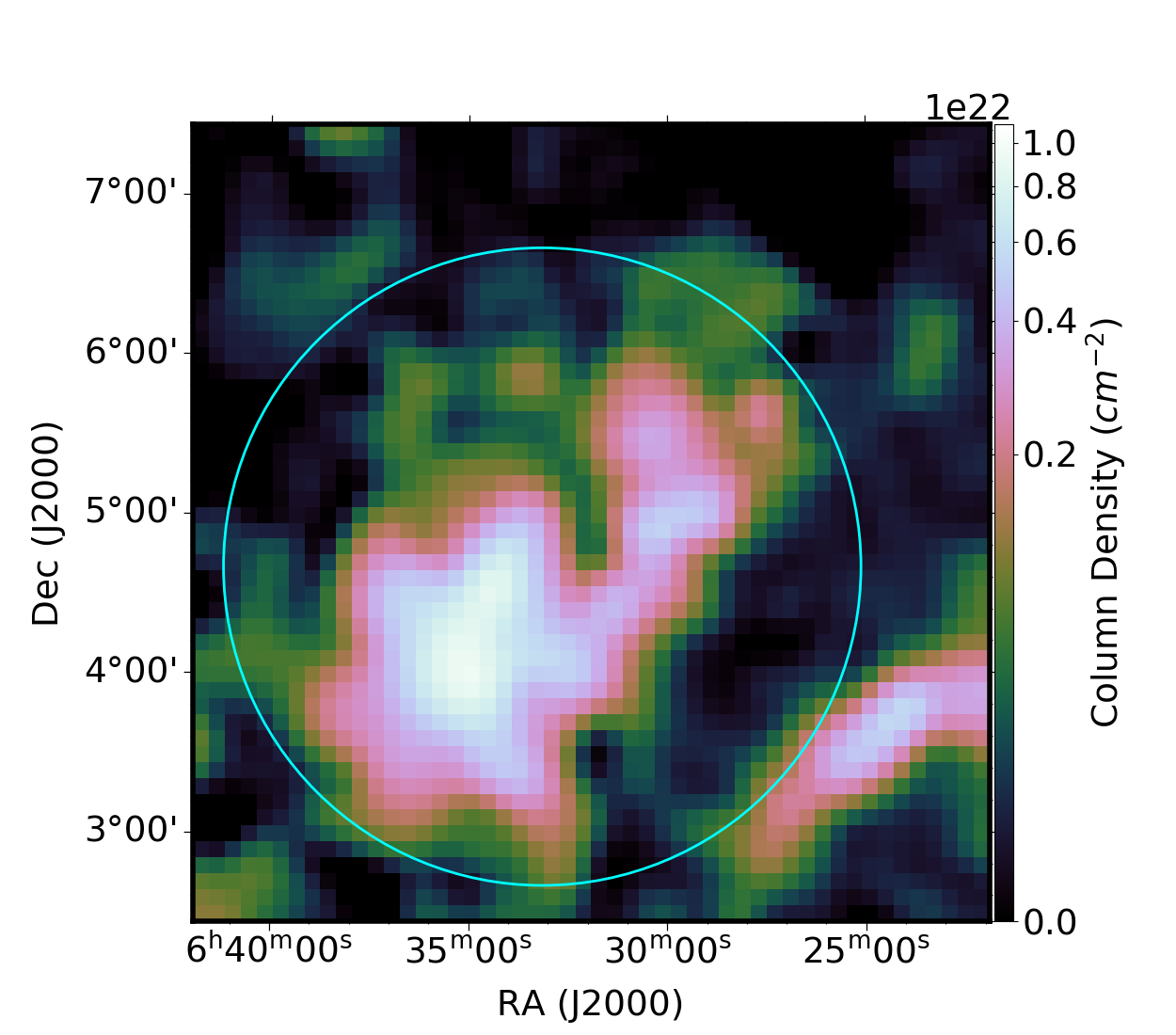}
        \label{res:1}
    \end{subfigure}
    \quad
    \begin{subfigure}[t]{0.4\linewidth}
        \centering
        \includegraphics[width=\linewidth]{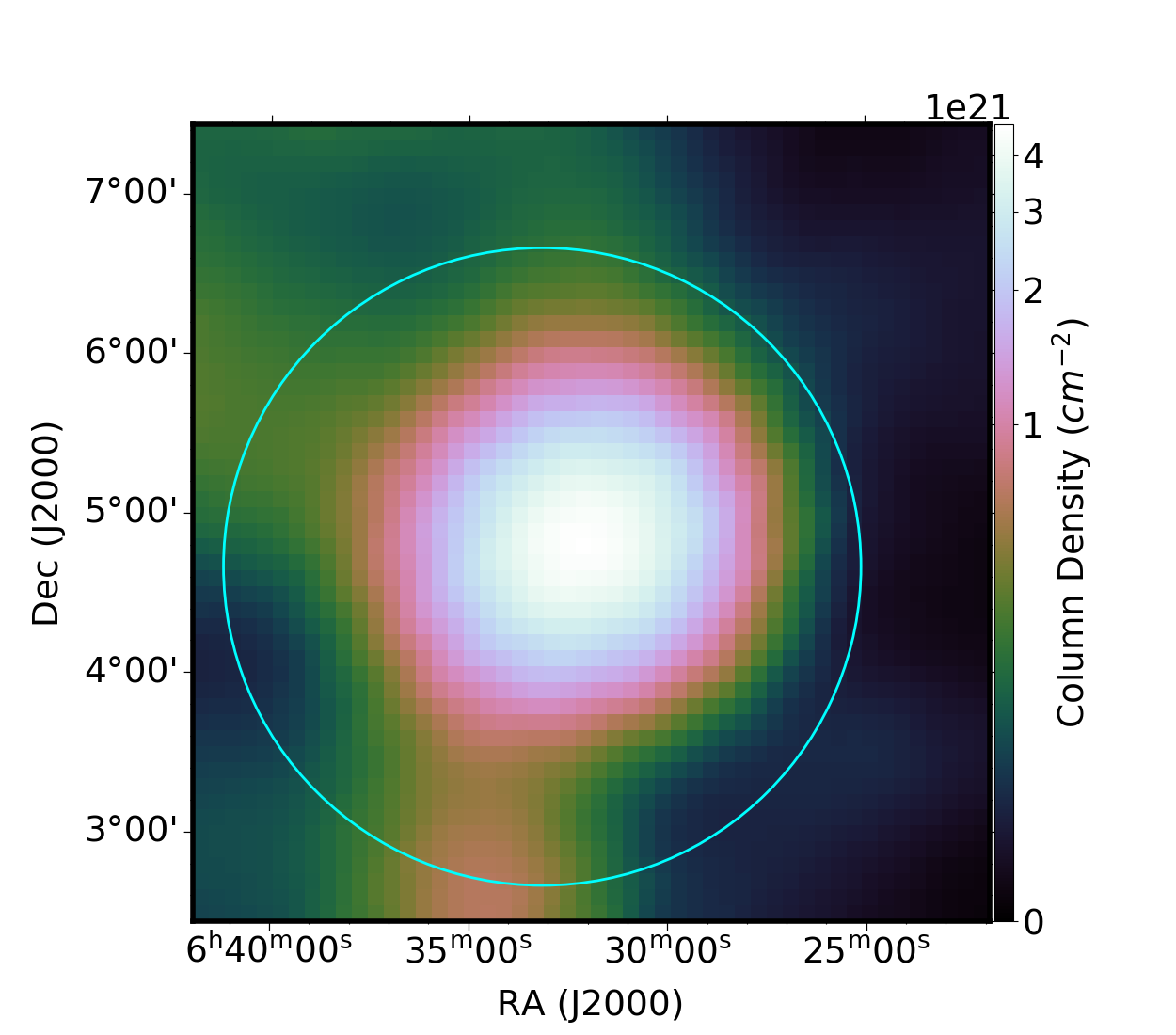}
        \label{res:2}
    \end{subfigure}
    \caption{Maps of gas column densities in two gas phases. Left shows the H$_{2}$ column density derived from the CO data \citep{CO}. Right shows the \ion{H}{ii} column density derived from the \planck\ free-free map assuming the effective density of electrons $n_{\rm e}=10~\rm cm^{-3}$. The regions selected to generate the spatial distribution templates are shown by the cyan circles. For details, see the context in Sec.\ref{sec:gas}.}\label{fig:gas} 
\end{figure*}

Both molecular gas (mainly H$_2$) and ionized gas (mainly \ion{H}{ii}) are detected in the vicinity of the Rosette Nebula. For the molecular gas, we used the composite CO survey data from \citet{CO} and derived the distribution of H$_2$, by assuming a linear relationship between the velocity-integrated brightness temperature of the CO 2.6-mm line, W$_\mathrm{CO}$, and the column density of molecular hydrogen, N(H$_2$). As proposed by  \cite{Lebrun1983}, $N\left(\mathrm{H}_2\right)$ can be obtained as $X_{\mathrm{CO}} \times W_{\mathrm{CO}}$, where the conversion factor $X_{\mathrm{CO}}=2.0\times10^{20}~$cm$^{-2}$K$^{-1}$km$^{-1}$s \citep{CO,Bolatto13}. Our CO template is the same as the work of \citet{Monoceros}, and we integrated the data over velocities of 0~km $\mathrm{s}^{-1}$ < V < 20~km $\mathrm{s}^{-1}$ based on the distance to the Rosette Nebula from Earth. The derived column density map of molecular gas is shown in the left panel of Fig.\ref{fig:gas}. The data in a circular region near the center of Rosette with obvious flux excess was retained to construct a spatial distribution template (hereafter referred to as the CO template), which is shown by the cyan circle in the left panel of Fig.\ref{fig:gas}. The region we selected was slightly different from the model generated by \citet{Monoceros}.  Given the distribution of column density, the total mass of this component was calculated to be $1.2\times 10^5~\rm M_\odot$. 

As for the \ion{H}{ii} column density, we used the \textit{Planck} free-free map \citep{Planck16} and applied the conversion factor from \cite{Finkbeiner03} to transform the emission measure into free-free intensity. And the \ion{H}{ii} column density is derived  via Eq.(5) from \cite{Sodroski97}:
\begin{equation}
\begin{aligned}
  N_{\ion{H}{ii}} = &1.2 \times 10^{15}\ {\rm cm^{-2}} \left(\frac{T_{\rm e}}{1\ \rm K}\right)^{0.35} \left(\frac{\nu}{1\ \rm GHz}\right)^{0.1}\left(\frac{n_{\rm e}}{1\ \rm cm^{-3}}\right)^{-1} \\
&\times \frac{I_{\nu}}{1\ \rm Jy\ sr^{-1}},
\end{aligned}
\end{equation}

where $I_{\nu}$ is the intensity of free-free emission, $\nu = \rm 353\ GHz$ is the frequency, and a electron temperature of $T_{e} =\rm 8000\ K$ is applied. For $n_{\rm e}$, an effective density $10\ \rm cm^{-3}$ is adopted, which is the value suggested in \cite{Sodroski97}. The derived \ion{H}{ii} column density map is shown in the right panel of Fig.\ref{fig:gas}. For the spatial distribution template of this component (hereafter referred to as the \ion{H}{ii} template), we chose the data in a circular region same as the one we chose for molecular gas, which is shown by the cyan circle in the right panel of Fig.\ref{fig:gas}. The total mass of this component was calculated to be $7.3\times10^4 ~\rm M_\odot$.%72676 
\label{sec:gas}

%%%%%%%%%%%%%%%%% data reduction and analysis %%%%%%%%%%%%%%%%%%
\section{\fermi\ data analysis}
\label{sec:data}
To study the gamma-ray emission towards the Rosette Nebula, we used the latest \fermi\ Pass 8 data from August 4, 2008 (MET 239557417) until March 3, 2022 (MET 667979611), and used the Fermitools from Conda distribution\footnote{\url{https://github.com/fermi-lat/Fermitools-conda/}} for the data analysis.
In this work, we chose a 20\deg $\times$ 20\deg\ square region centered at the position of the Rosette Nebula (R.A. = 97.979$\deg$, Dec. = 4.942$\deg$) as the region of interest (ROI).
We used {\sl gtselect} to select "source" class events (evtype = 3 and evclass = 128) with zenith angles less than 90$^{\circ}$ to filter out the background contamination from the Earth's limb. We also selected the good time intervals by applying the recommended expression $\rm (DATA\_QUAL > 0) \&\& (LAT\_CONFIG == 1)$. The instrument response functions {\it P8R3\_SOURCE\_V3} were applied to analyze the SOURCE events. 
We performed a standard binned analysis following the official tutorial of binned likelihood analysis\footnote{\url{https://fermi.gsfc.nasa.gov/ssc/data/analysis/scitools/binned_likelihood_tutorial.html}}. We generated the source models using make4FGLxml.py\footnote{\url{https://fermi.gsfc.nasa.gov/ssc/data/analysis/user/make4FGLxml.py}}, which consists of the sources in the LAT 12-year Source Catalog  (4FGL-DR3, \citealt{4FGL}) those are within the ROI enlarged by 10$^{\circ}$, the Galactic diffuse background emission (gll\_iem\_v07.fits), and the isotropic emission background (iso\_P8R3\_SOURCE\_V3\_v1.txt). In addition, the spectral parameters of sources within 10$\deg$ from the center and the normalization factor of the Galactic diffuse background and isotropic background were set free.
\subsection{Spatial analysis}

\begin{figure*}
    \centering
    \includegraphics[width=0.5\linewidth]{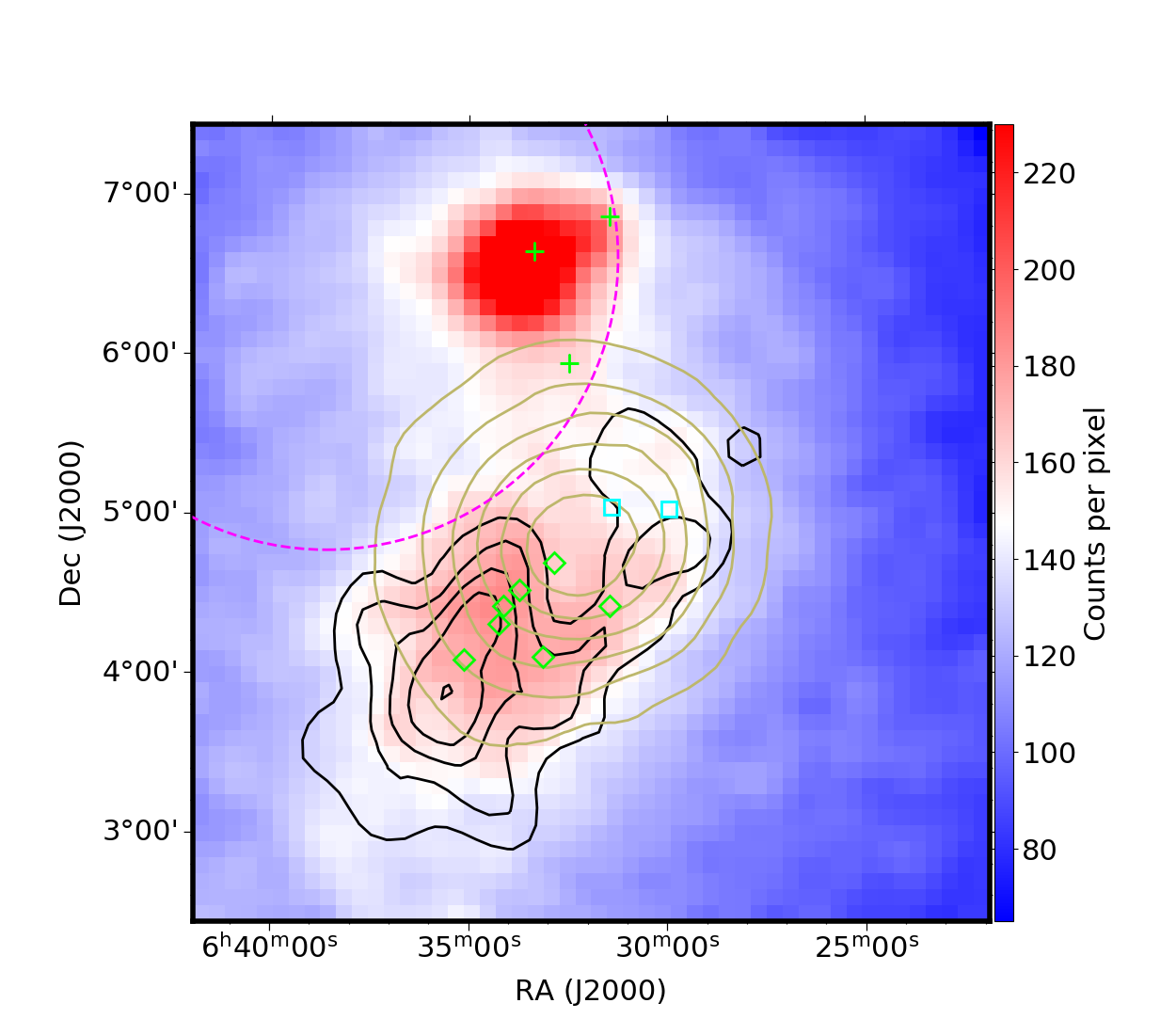}
    \caption{Gamma-ray counts map of around the Rosette Nebula. The magenta dashed circle shows the position of the Monoceros Loop. The point-like sources in 4FGL-DR3 are shown in green crosses. The green diamonds represent the positions of seven embedded clusters in the RMC identified by \citet{1997clusters}. The cyan box shows the position of NGC 2237 and NGC 2244 \citep{2008NGC2237}. The black contours show the spatial distribution of the H$_{2}$ gas while the brown contours show the distribution of the \ion{H}{ii} gas.}   
\label{fig:cmap}
\end{figure*}

First, we used the events from 0.1 to 500~GeV to study the spatial distribution of the gamma-ray emission near the nebula. The gamma-ray counts map in the $5\deg \times 5\deg$ region around the Rosette Nebula is shown in Fig.\ref{fig:cmap}. 
In the 4FGL-DR3 catalog, the spatial models of Rosette Nebula and Monoceros Loop are adopted from the work of \cite{Monoceros}, in which the nebula represented by a template derived from the CO gas distribution and the SNR is fitted as a Gaussian disk. Besides, there are three point-like sources in 4FGL-DR3 overlapping with the Rosette Nebula along the line of sight, each with a corresponding identified source (4FGL J0632.8+0550 and HESS J0632+057, 4FGL J0633.7+0632 and PSR J0633+0632, 4FGL J0631.8+0645 and PSR J0631+0646).

\begin{figure*}
    \centering
    \begin{subfigure}[t]{0.4\linewidth}
        \centering
        \includegraphics[width=\linewidth]{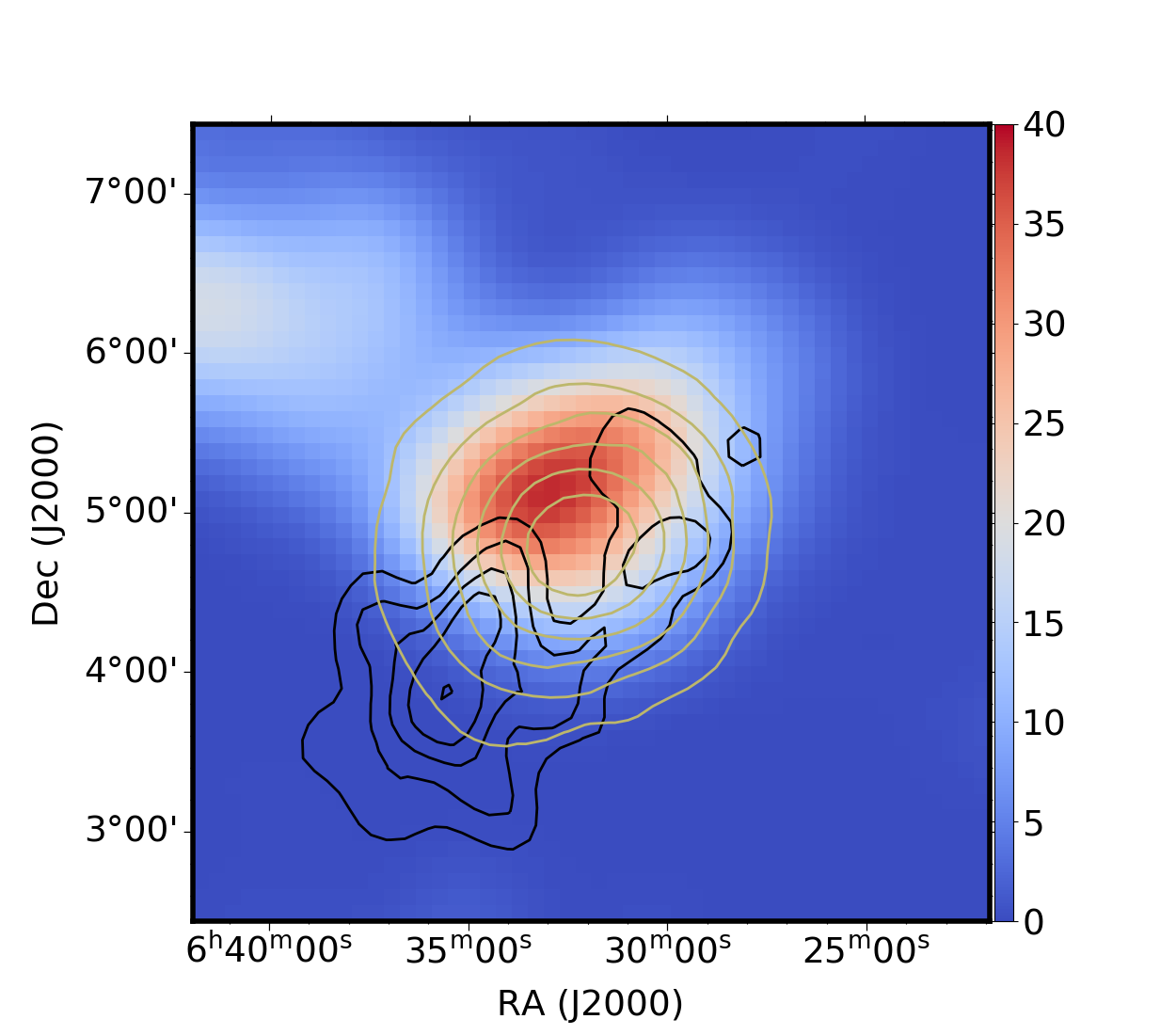}
        \caption{CO model}\label{ts:1}
    \end{subfigure}
    \quad
    \begin{subfigure}[t]{0.4\linewidth}
        \centering
        \includegraphics[width=\linewidth]{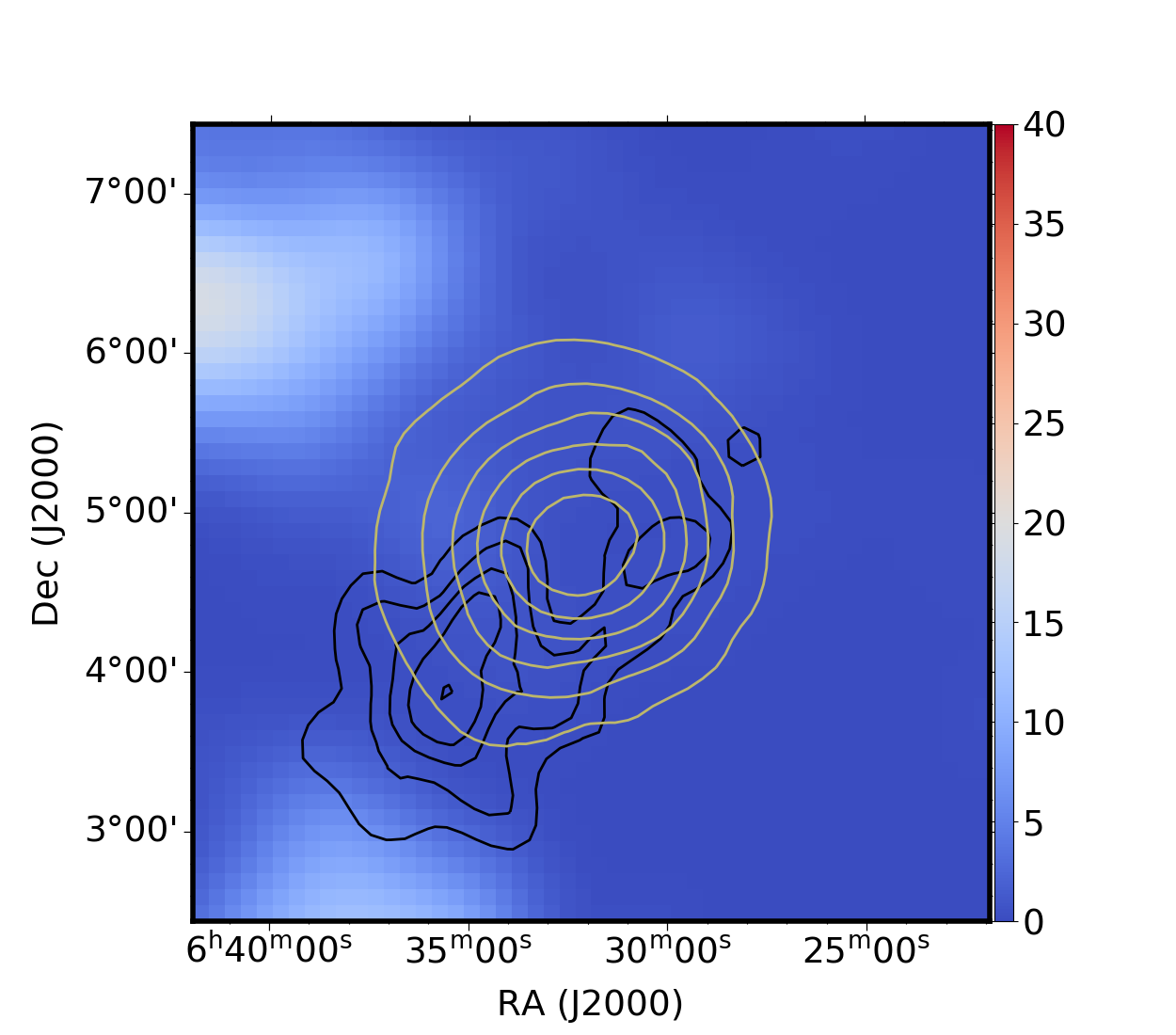}
        \caption{CO + \ion{H}{ii} model}\label{ts:2}
    \end{subfigure}
    \caption{Residual gamma-ray TS maps obtained from different source models for energy range 0.1-500~GeV, smoothed with Gaussian filter of 0.3$^{\circ}$. The black and brown contours show the spatial distribution of the CO template and the \ion{H}{ii} template, respectively.}\label{fig:ts} 
\end{figure*}

With the sources in the 4FGL-DR3 catalog, we conducted a binned likelihood analysis and applied the fitted 4FGL-DR3 model to calculate the residual test statistic (TS) map around the nebula. The TS value is defined as $2\left(\log \mathcal{L} / \mathcal{L}_0\right)$, in which $\mathcal{L}_0$ is the likelihood of null hypothesis and $\mathcal{L}$ is the likelihood with a test source included at a given position. However, evident {excess around the nebula was found from the residual TS map}, indicating a more detailed analysis is required to study the gamma-ray emission from the Rosette Nebula. We then replaced the Rosette template in the 4FGL-DR3 model with the CO template which is described in Sec.\ref{sec:gas}, and applied the log-parabolic spectral function for the CO template, which is the same setting for the Rosette template in the 4FGL-DR3 catalog (hereafter referred to as the CO model). The analysis results of the CO model are basically the same as that of the original 4FGL-DR3 model since both the CO template and the Rosette template in 4FGL-DR3 are derived from the same CO data \citep{Monoceros}. Both models obtained similar -$\log({\cal L})$ of about -2453860 and their residual TS maps show a similar distribution of the excess emission around the nebula.  From the residual TS map of the CO model (as shown in the left panel of Fig.\ref{fig:ts}), we can see the distribution of residual gamma rays mainly overlaps with the distribution of the \ion{H}{ii} gas.

\begin{table*}
\caption{\fermi\  data fitting results of different models}  
\label{tab:likelihood}       
%\centering                        
\begin{tabular}{cccc}       
\hline\hline     
Model&Free Parameters (k)&-$\log({\cal L})$&AIC\\
\hline
CO model&77&-2453860&-4907566\\
%& \ion{H}{ii}&$9.49\pm3.14$ & $2.58\pm0.22$\\
CO+\ion{H}{ii} model&80&-2453911&-4907662\\
%& \ion{H}{ii}&$9.49\pm3.14$ & $2.58\pm0.22$\\
Gaussian model&80&-2453858&-4907556\\
\hline
\end{tabular}
\end{table*}

As we studied the \ion{H}{ii} density in the same region in Sec.\ref{sec:gas}, the column density of \ion{H}{ii} was found to be of the same order of magnitudes with molecular gas, and with quite a different distribution seemingly related to the excess in the residual TS map obtained using the CO model (see the left panel of Fig.\ref{fig:ts}). Thus, we added a new source of which the spatial model is the \ion{H}{ii} template and the spectral function is log-parabola to the CO model (hereafter referred to as CO+\ion{H}{ii} model), and performed a new likelihood analysis. As shown in the TS map derived from the CO+\ion{H}{ii} model (the right panel of Fig.\ref{fig:ts}), the residual emission is negligible after adding the \ion{H}{ii} component. 
Moreover, we also tried to use a radial Gaussian disk with the spectral function of log-parabola to represent the residual emission in the left panel of Fig.\ref{fig:ts}. The disk was centered at the TS peak position (R.A. = 98.531$\deg$, Dec. = 4.992$\deg$), with a sigma ranging from 0.1$^\circ$ to 0.9$^\circ$. The best-fit result is the disk with a sigma of 0.2 $^\circ$ (hereafter referred to as the Gaussian model).
To find which model fits the data best, we applied the Akaike information criterion (AIC). AIC is calculated as $-2 \log ($ likelihood $)+2 \mathrm{k}$, in which k is the number of free
parameters in the model. The -$\log({\cal L})$ and corresponding AIC value of the models are listed in Table\ref{tab:likelihood}. As shown in Table.\ref{tab:likelihood}, the best fit for the data is the CO+\ion{H}{ii} model.

\subsection{Spectral analysis}
\label{sec:spectral_analy}
\begin{figure*}
\label{fig:gam}
    \centering
    \begin{subfigure}[t]{0.4\linewidth}
        \centering
        \includegraphics[width=\linewidth]{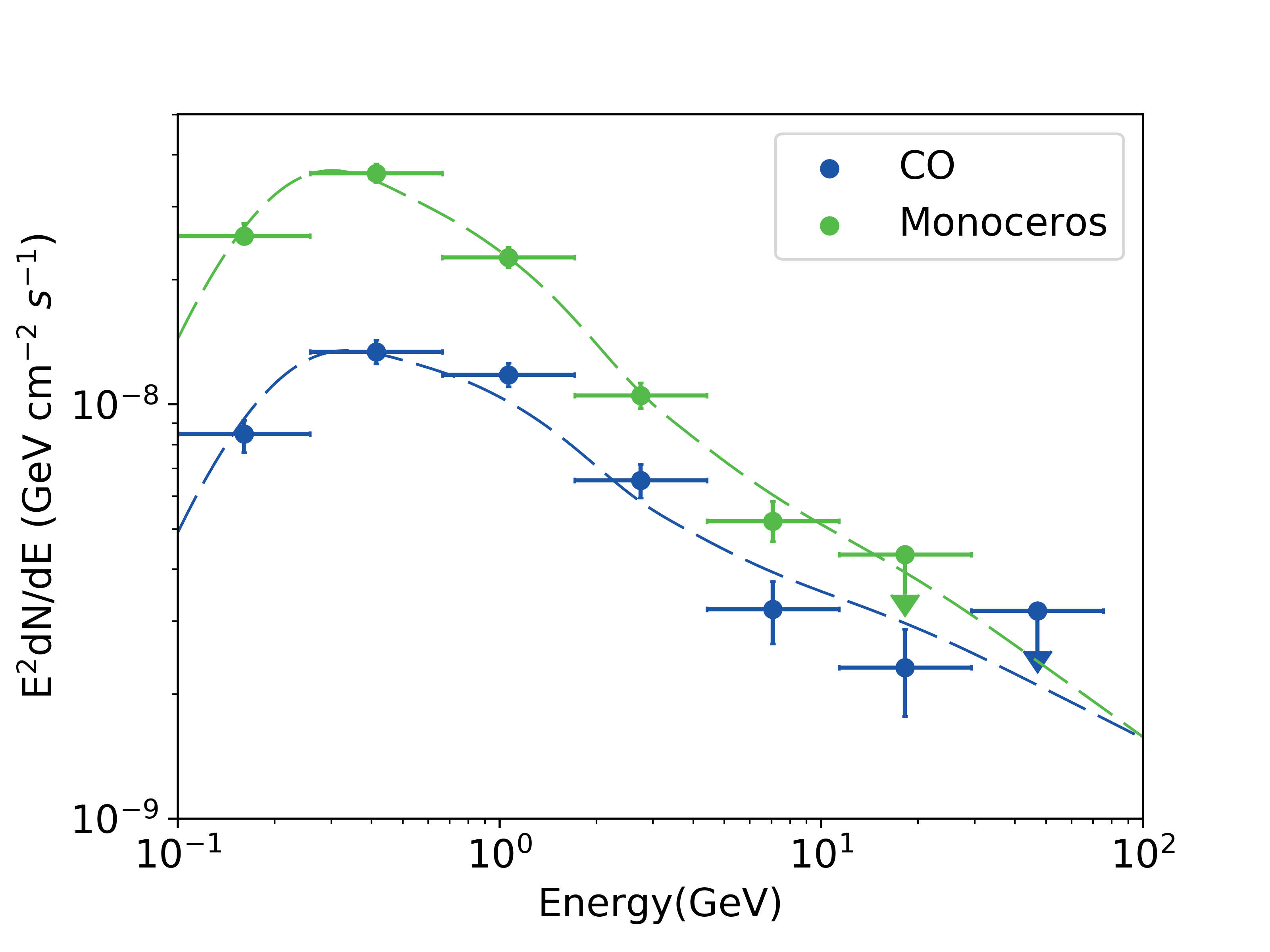}
        \caption{CO model}\label{gamma:1}
    \end{subfigure}
    \quad
        \begin{subfigure}[t]{0.4\linewidth}
        \centering
        \includegraphics[width=\linewidth]{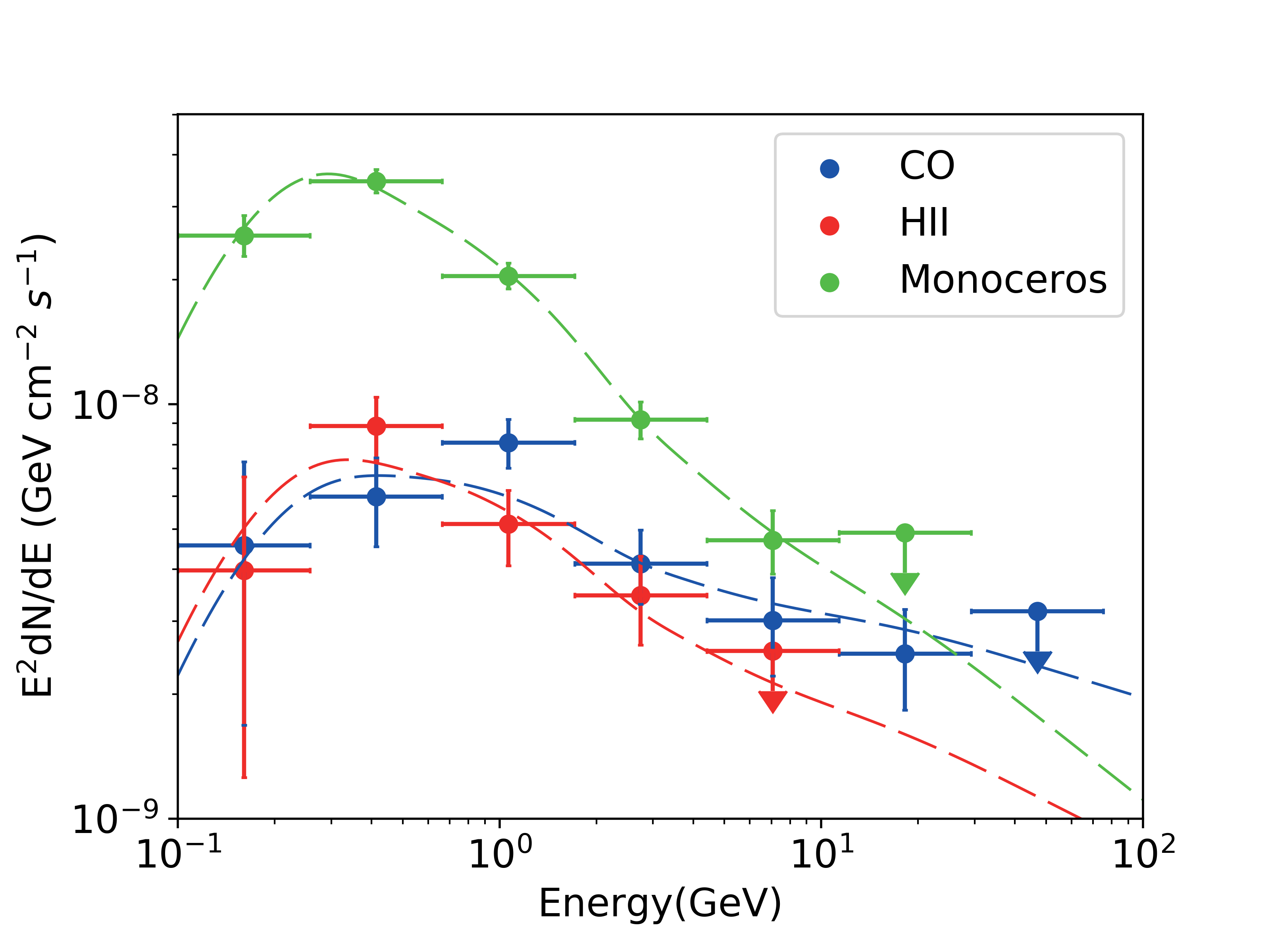}
        \caption{CO+\ion{H}{ii} model}\label{gamma:2}
    \end{subfigure}
    \caption{Gamma-ray spectral energy distributions of the Rosette Nebula and the Monoceros SNR obtained from different source models. For each data point, the error bar indicates the combined statistical and systematic uncertainty. Data points with TS values lower than 4 are replaced by their upper limits. The dashed lines show the corresponding pion-decay emission derived from the best-fit CR spectra (see Sec.~\ref{sec:CR} for details.).} \label{fig:gamma} 
\end{figure*}

Based on the above spatial analysis, we derived the spectral energy distribution (SED) of the Rosette Nebula of the CO model and the  CO+\ion{H}{ii} model respectively. Here, we divided the data into nine logarithmically spaced energy bins and obtained the SED by applying maximum-likelihood analysis for each bin. 
Systematic uncertainties of the flux for each energy bin were also estimated using the method described on the Fermi official website \footnote{\url{https://fermi.gsfc.nasa.gov/ssc/data/analysis/scitools/Aeff_Systematics.html}}. 
Besides,  we calculated the 99\% flux upper limit for energy bins with the source TS value less than 4. The derived SEDs of different components in different models are shown in Fig.\ref{fig:gamma}.

%----------------------------------------------------- FIGURE 3
  
\begin{figure*}
\label{figure:cr}
    \centering
    \begin{subfigure}[t]{0.4\linewidth}
        \centering
        \includegraphics[width=\linewidth]{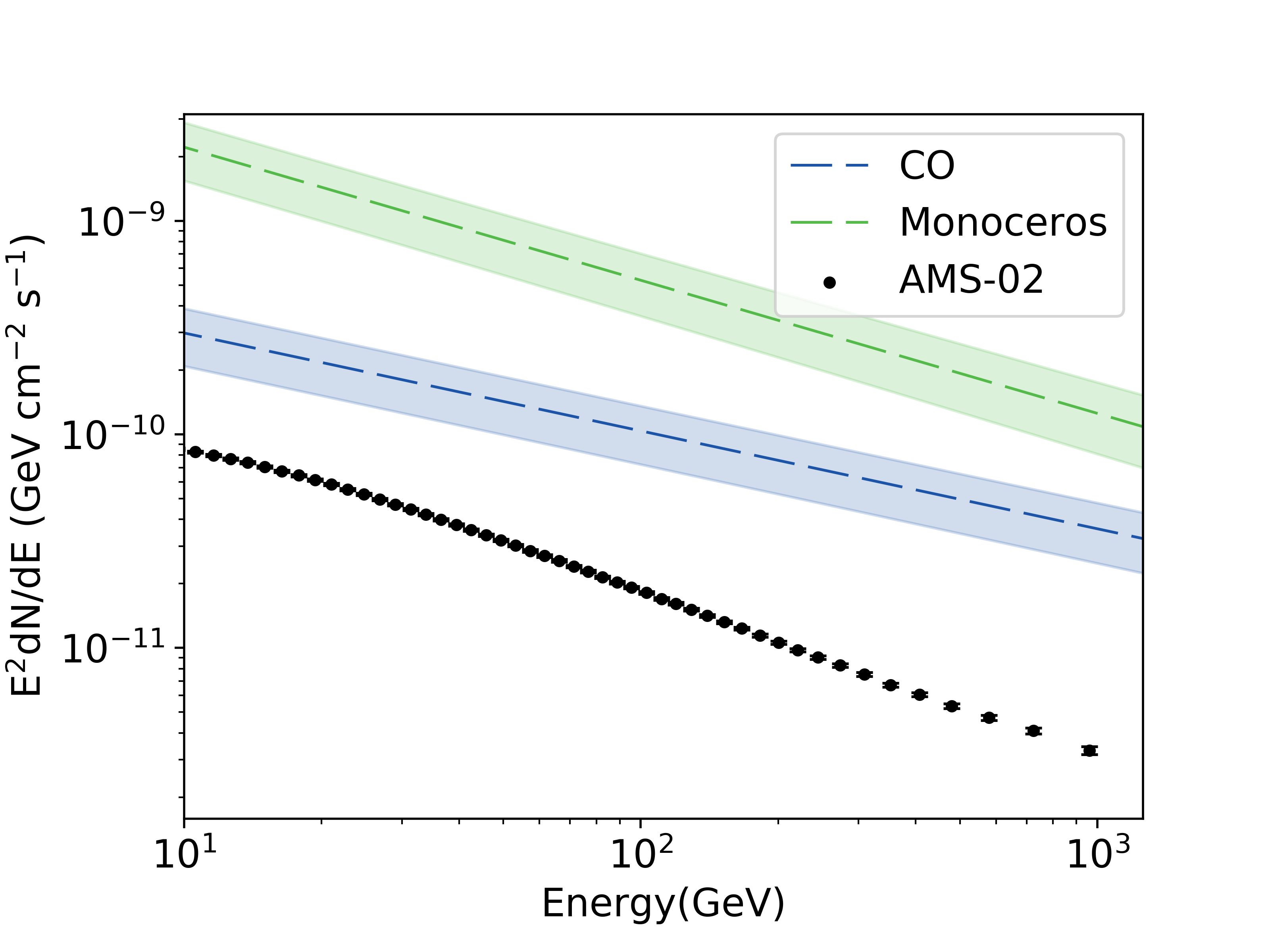}
        \caption{CO model}\label{cr:1}
    \end{subfigure}
    \quad
        \begin{subfigure}[t]{0.4\linewidth}
        \centering
        \includegraphics[width=\linewidth]{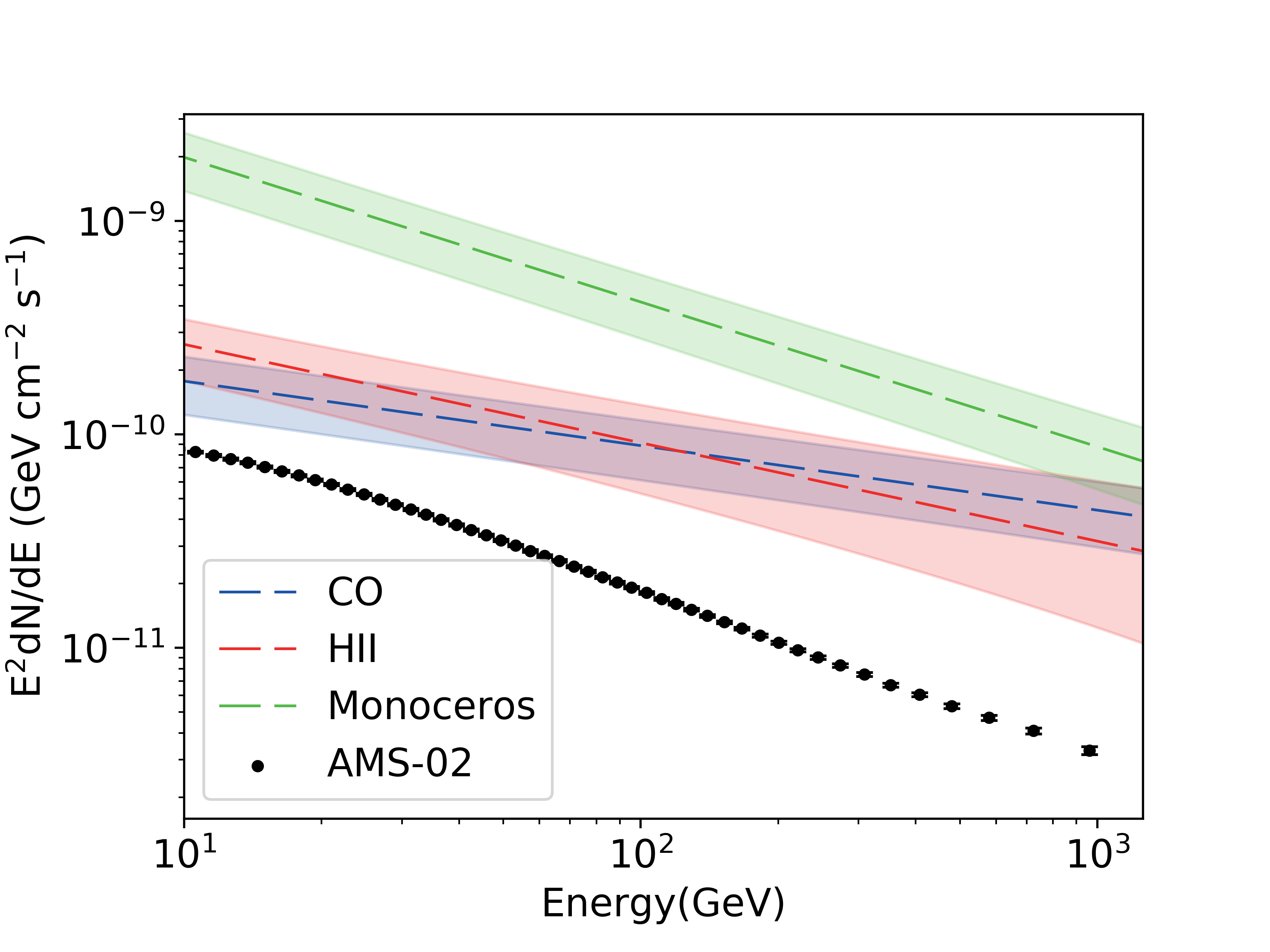}
        \caption{CO+\ion{H}{ii} model}\label{cr:2}
    \end{subfigure}
    \caption{Derived CR spectra of the Rosette Nebula and the Monoceros SNR for different models. The dashed lines show the best-fit CR spectra and the shaded regions represent the 1 $\sigma$ errors including both the statistics and the systematic uncertainty. The black dots are the AMS-02 observation data acquired from \citet{Aguilar15}.}\label{fig:cr} 
\end{figure*}

\begin{table*}
\caption{Fitted spectral parameters of CR proton for different gas contents of the Rosette Nebula and the Monoceros SNR}  
\label{tab:cr}       
%\centering                        
\begin{tabular}{cccc}       
\hline\hline     
Model&Component&$f_i(10^{-10}\,\mathrm{cm}^{-2}\, \mathrm{s}^{-1}\,\mathrm{GeV}^{-1})$&$\Gamma$\\
\hline
\multirow{2}{*}{CO model}&CO&$8.59\pm0.35$&$-2.46\pm0.02$\\
&Monoceros&$93.17\pm6.23$  & $-2.62\pm0.04$\\
\hline
\multirow{3}{*}{CO+\ion{H}{ii} model}&CO&$3.57\pm0.30 $&$-2.30\pm0.03$\\
&\ion{H}{ii}&$7.63\pm1.24$  & $-2.46\pm0.10$\\
&Monoceros&$94.72\pm10.04$  & $-2.68\pm0.04$\\
\hline
\end{tabular}
\end{table*}

\section{CR content in the vicinity of the Rosette Nebula}
\label{sec:CR}
We assumed that the gamma rays are mainly generated from the proton-proton interaction between CR protons and the gases, which enabled us to derive the spectra of CR in the vicinity of the Rosette Nebula using the information of the gamma-ray spectra as well as the derived gas masses. 
We used the gamma-ray production cross section from \citet{2014pp} and assumed a power-law proton spectrum $F=f_iE^{\Gamma}$ to perform a likelihood fitting to the gamma-ray data.  
Here $F$ is the flux density of protons in unit of $\mathrm{cm}^{-2}\, \mathrm{s}^{-1}\,\mathrm{GeV}^{-1}$,  $E$ is the proton energy in unit of GeV, and $\Gamma$ stands for the spectral index. Given the gas masses obtained in Sec.\ref{sec:gas}, we derived the absolute CR fluxes associated with the CO and \ion{H}{ii} components, respectively.  Especially, for Monoceros, we used an average gas density of 3.6~cm$^{-3}$ based on the \ion{H}{i} observations \citep{Mon_density}. The best-fit results are listed in Table~\ref{tab:cr} and illustrated in Fig.~\ref{fig:cr}. The spectral index of CR protons associated with the CO gas is a bit harder after adding the \ion{H}{ii} template and is much harder than that of the Monoceros SNR. Besides, the dashed lines in Fig.\ref{fig:gamma} represent the corresponding gamma-ray emissions derived from the CR fluxes and gas masses. As shown in Fig.~\ref{fig:cr}, the fluxes of CR protons that are associated with the nebula and the SNR are higher than the local CR proton flux measured by AMS-02 \citep{Aguilar15}.

\section{Discussion and conclusion}
 
\label{sec:conc}

In this paper, we analyzed {\sl Fermi}-LAT data towards the Rosette Nebula with a time span of more than 13 years. We found that the extended gamma-ray emission in the vicinity of the nebula can be well modeled by the CO gas distributions, which is consistent with the former results \citep{Monoceros}. However, We found that adding the \ion{H}{ii} gas template can significantly improve the fitting results.

After including this \ion{H}{ii} gas template, the derived gamma-ray emissions associated with the Rosette Nebula (CO template and \ion{H}{ii} template) and the SNR Monoceros show clear low energy break below $1~\rm GeV$, which are strong hints that the gamma-ray emissions from both regions are from the pion-decay process in the interaction of CR protons with ambient gas. Such results are also similar to that in the former study \citep{Monoceros}. However, the gamma-ray emission from the Rosette Nebula is now significantly harder than that from Monoceros. And the CR associated with the ionized gas has a similar index (about $-2.46$) to that associated with the molecular gas (about $-2.30$), harder than that of the Monoceros SNR (from $-2.6$ to $-2.7$). This is substantially different from the results in the former study in \citet{Monoceros}, in which the similar spectral shape in both regions is used implying that the gamma-ray emissions from both regions are from the same CR population accelerated by the SNR Monoceros.

The different gamma-ray spectra in Rosette and Monoceros cannot formally rule out the possibility that the gamma-ray emissions in the nebula are produced by the CRs that are accelerated and injected by the SNR. The harder spectrum in the nebula may be formed by propagation effects, that is, due to the energy-dependent diffusion only the high-energy CRs reached the Rosette Nebula. Assuming the Monoceros SNR and the Rosette Nebula are located at the same distance (1.55 kpc), the distance between the center of Monoceros SNR and Rosette Nebula is about $60~\rm pc$. In case the harder spectrum of the Rosette Nebula is attributed to that the CR below about $10~\rm GeV$ has not reached Rosette Nebula yet and taken into account the age of the SNR of $3\times 10^4$ years \citep{Monoceros}, the required diffusion coefficient is about $10^{28}~\rm cm^2/s$ at $10~\rm GeV$, which is about one order of magnitude smaller than the value in the interstellar medium \citep{strong98}. Such a low diffusion coefficient is also observed in other SNR-MC systems \citep[e.g.,][]{Fujita2010,lihui2012,liu2015kes41}, which may be caused by a higher turbulence level near the CR accelerators. However, if the distance between the SNR and the nebula is $\sim 0.4$\,kpc as the recent measurement made by \citet{2018zhao},  the diffusion coefficient is about  $10^{30}~\rm cm^2/s$ at 10 GeV for the protons accelerated by the SNR to reach the nebula, which is about ten times larger than the value in the interstellar medium.

On the other hand, it is also possible that the gamma rays from the Rosette Nebula are produced by CRs other than those accelerated by Monoceros SNR. The recent estimation of the distance of the Rosette Nebula also suggests that it doesn't interact with the SNR \citep{2018zhao}. A natural acceleration site is the star cluster NGC 2244 inside the Rosette Nebula. NGC 2244 is a YMC with an age of about $2~\rm Myr$ \citep{muzic22}, and harbors more than 20 stars with a spectral type early than B3 \citep{wang2009}. The total wind power can be estimated to be at the order of  $10^{37} ~\rm erg/s$. Assuming an acceleration efficiency of 10\%, the total CR accelerated in NGC 2244 can be estimated as $10^{50}~\rm erg$. The gamma-ray luminosity in the Rosette Nebula is about $10^{34}~\rm erg/s$, taken into account of the gas density of about $10~\rm cm^{-3}$, and the total energy in the CR content is $\sim 10^{49} ~\rm erg$. Therefore, NGC 2244 is powerful enough to account for the gamma-ray emissions in the Rosette Nebula. Furthermore, the hard gamma-ray spectrum is also similar to the gamma-ray emission in other YMC systems, such as Cygnus cocoon \citep{Ackermann11, Aharonian19}, NGC 3603 \citep{Yang17}, W40 \citep{sunw40}, W43 \citep{yang20w43}, NGC 6618 \citep{liub2022m17}.
    
In conclusion, we revisited the GeV gamma-ray emission towards the Rosette Nebula and found the inclusion of an \ion{H}{ii} gas template can significantly improve the likelihood of the fitting results. With the new spatial template, we found the gamma-ray emissions from the Rosette Nebula are substantially harder than previously reported. Although we cannot rule out the possibility that the gamma-ray emission from this region is produced by CRs accelerated by the nearby SNR Monoceros. A more natural explanation is that the CR accelerated by YMC NGC 2244 illuminated both molecular and ionized gases in the Rosette Nebula. In this case, the Rosette Nebula is another example of the gamma-ray-emitting YMC system.

\section{Acknowledgements}

Rui-zhi Yang is supported by the NSFC under grant 12041305 and the national youth thousand talents program in China. Bing Liu acknowledges the support from the NSFC under grant 12103049.

\section{Data availability}
The \fermi\ data used in this work is publicly available, which is provided online by the NASA-GSFC Fermi Science Support Center\footnote{\url{ https://fermi.gsfc.nasa.gov/ssc/data/access/lat/}}.
We make use of the CO data\footnote{\url{ https://lambda.gsfc.nasa.gov/product/}} to derive the H$_{2}$. The data from \planck\ legacy archive\footnote{\url{ http://pla.esac.esa.int/pla/\#home}} are used to derive the \ion{H}{ii} gas content.
The \ion{H}{i} data are from the HI4PI\footnote{\url{http://cdsarc.u-strasbg.fr/viz-bin/qcat?J/A+A/594/A116}}.

\bibliographystyle{mnras}

\bibliography{ms}

\bsp	% typesetting comment
\label{lastpage}
\end{document}